\documentclass[
showpacs,preprintnumbers,amsmath,amssymb,11pt]{revtex4}
\usepackage{amsmath}
\usepackage{graphicx}

\begin{document}

\tolerance=5000

\title{Reconstructing cosmic acceleration from modified and non-minimal
gravity:  The Yang-Mills case}

\author{E. Elizalde and A.~J. L\'opez-Revelles}

\affiliation{Consejo Superior de Investigaciones Cient\'{\i}ficas, ICE/CSIC-IEEC, Campus UAB, Facultat de
Ci\`encies, Torre C5-Parell-2a pl, E-08193 Bellaterra (Barcelona) Spain}

\begin{abstract}
A variant of the accelerating cosmology reconstruction program is developed for $f(R)$ gravity and for a modified Yang-Mills/Maxwell theory. Reconstruction schemes in terms of e-foldings and by using an auxiliary scalar field are developed and carefully compared, for the case of $f(R)$ gravity. An example of a model with a transient phantom behavior without real matter is explicitly discussed in both schemes. Further, the two reconstruction schemes are applied to the more physically interesting case of a Yang-Mills/Maxwell theory, again with explicit examples. Detailed comparison of the two schemes of reconstruction is presented also for this theory. It seems to support, as well, physical non-equivalence of the two frames.
\end{abstract}


\maketitle


\section{INTRODUCTION}

    One of the most important problems of modern cosmology (and theoretical physics too) is the explanation of the current universe speed-up, first discovered in \cite{4,5}. A convenient way to express this situation is to introduce a new form of energy, called Dark Energy (DE). Indeed, it is not easy to generate the necessary amount of repulsive force (not less the very fact that it is repulsive) through quantum vacuum fluctuation contributions of ordinary fields, at cosmological scale \cite{eli1}. To start with, one needs to find an acceptable solution to the cosmological constant problem. During the last few years several theories, which are extensions of or alternatives to Einstenian Gravity, have been developed in order to formulate and try to explain the dark energy universe. The most accurate observational data we now have indicate that the equation of state (EoS) parameter, $\omega$, for DE is very close to $-1$ (for a review of observational data from the theoretical point of view, see e.g.~\cite{8}, and for a description of the observable cosmological parameters, see e.g.~\cite{9}).

    As gravitational alternatives for DE, modified gravity theories have been formulated, calling for plausible late-time modification of General Relativity (GR). Many modified gravity models have been proposed in the literature (for a review, see \cite{1}), starting from the very simple $1/R$ theory \cite{16,10,11} (that soon was declared as problematic) to more elaborated ones which, being still not fundamental, are already quite often inspired by string and M-theory considerations \cite{12,13}.

    There are some modified gravity theories (see e.g.~\cite{16,7}) which have the capacity to achieve a reasonably natural unification of the early-time inflation epoch with the late-time acceleration period, in which we are now, by taking advantage of the different role played by the gravitational terms relevant at small and at large curvatures. Moreover, if one assumes that the universe may enter in a phantom phase ($\omega < -1$) \cite{kom1,25}, modified gravity may naturally describe the transition from the ordinary, non-phantom phase towards the phantom one without the necessity to introduce exotic matter. Also, no Big Rip is usually expected there, because in modified gravity the phantom phase is often transient. These theories may also be able to encompass, in a truly unified way, DE and dark matter.

    For any such theory to be valid it is always strictly required that it accurately describes the known sequence of cosmological epochs, specifically it must fit very well an increasing number of more and more precise observational data \cite{14,15}. In particular, it must pass the Solar System tests (for $f(R)$ models, this is discussed e.g.~in Refs.~\cite{16,17,18,19}, while the case of SdS metrics is studied in \cite{20}). After all this procedure, it turns out, in the end, that the possibility exists to pass all known local tests \cite{6} and to construct a realistic $f(R)$ modified gravity theory which unifies early-time inflation with late-time acceleration \cite{7}. As a consequence, a natural resolution of the problem at hand seems to be on the right track.

    Another important issue to be taken into account concerns the appearance of different types of finite-time, future cosmological singularities (see \cite{29}), what is a typical property of the number of modified gravities with the effective quintessence/phantom behavior, in the same way as it had been found for other, more simple scalar/fluid dark energy models. It has been proven (see \cite{32, 33}) that in modified gravities all four known types of finite-time future singularities may possibly appear. However, modified gravity has the chance to cure all of these future singularities by the addition of an $R^2$-term which is relevant at the very early stages of the universe only (see \cite{31}). The combination of non-singular realistic modified gravity with singular realistic models unifying the early-time inflation epoch with late-time acceleration finally results in a realistic, non-singular theory (for more details see \cite{30}). For all these reasons, modified gravity theories constitute a very attractive alternative to dark energy.

    In this paper, by making one further step in the direction of trying to build a truly realistic theory, we develop a non-trivial variant of the accelerating cosmology reconstruction program both for $f(R)$ gravity and for a modified Yang-Mills/Maxwell theory (see \cite{relw} for related work). In Sect.~II, two seemingly different reconstruction schemes, (i) in terms of e-foldings (for a general review, see \cite{21}), and (ii) by using an auxiliary scalar field (see, e.g.~\cite{22,23,24,25}) are reviewed and then explicitly compared, what is done here for the first time. To illustrate the results, the example of a model with a transient phantom behavior without real matter is discussed in both schemes. In Sect.~III the corresponding reconstruction schemes for the case of a Yang-Mills theory (in a extended, non-minimal version) are developed, one of them using an auxiliary scalar field and the other one without it. The specific example of power-law expansion is carefully considered in both schemes. Finally, in the last section, a summary of the results obtained and, in particular, of the detailed comparison with pros and cons of the two schemes of reconstruction is presented.

\section{COSMOLOGICAL RECONSTRUCTION OF REALISTIC MODIFIED $f(R)$ GRAVITIES}

    \subsection{GENERAL FORMULATION IN TERMS OF E-FOLDINGS}

        We will shaw in this section how one can construct an $f(R)$ model realizing a given cosmology, by using the techniques of \cite{2}. For the benefit of the reader (and self-consistency of the paper), in a simpler situation, and with the help of an explicit example, we shall review the method that will be later applied to the Yang-Mills case. The starting action for $f(R)$ gravity (see e.g.~\cite{1}, for a general review) is
            \begin{equation}\label{t1}
                        S = \int d^4 x \sqrt{-g} \left( \frac{f(R)}{2 \kappa^2} + \mathcal{L}_{matter} \right).
            \end{equation}

        The first FRW equation turns into the following field equation
            \begin{equation}\label{t2}
                0 = -\frac{f(R)}{2} + 3 \left( H^2 + \dot H \right) f'(R) - 18 \left( 4 H^2 \dot H + H \ddot H \right) f''(R) + \kappa^2 \rho,
            \end{equation}
        with $R = 6 \dot H + 12 H^2$. Using the e-folding variable, $N = \ln{\frac{a}{a_0}}$, instead of the cosmological time $t$,  one gets
            \begin{equation}\label{t3}
                0 = - \frac{f(R)}{2} + 3 \left( H^2 + H H'\right) f'(R) - 18 \left( 4 H^3 H' + H^2 (H')^2 + H^3 H'' \right) f''(R) + \kappa^2 \rho,
            \end{equation}
        where $H' \equiv \frac{dH}{dN}$. Assuming the matter density $\rho$ is given in terms of a sum of fluid densities with constant EoS parameters, $\omega_i$, we have
            \begin{equation}\label{t4}
                \rho = \sum \limits_i \rho_{i0} a^{-3 (1 + \omega_i)} = \sum \limits_i \rho_{i0} a_0^{-3 (1 + \omega_i)} e^{-3 (1 + \omega_i) N}.
            \end{equation}
        Using the Hubble rate $ H = g(N) = g(-\ln{(1 + z)})$,   with $z = e^{- N} - 1$ the redshift, the scalar curvature takes the form: $R = 6 g'(N) g(N) + 12 g(N)^2$, which can be solved with respect to $N$ as $N = N(R)$. Defining $G(N) \equiv g(N)^2 = H^2$ and using (\ref{t4}), Eq.~(\ref{t3}) yields
            \begin{equation}\label{t6}
                0 = - 9 G(N(R)) \left[ 4 G'(N(R)) + G''(N(R)) \right] \frac{d^2 f(R)}{dR^2} + \left[ 3 G(N(R)) + \frac{3}{2} G'(N(R)) \right] \frac{d f(R)}{dR} - \frac{f(R)}{2} +$$
                $$+ \sum \limits_i \rho_{i0} a_0^{-3 (1 + \omega_i)} e^{-3 (1 + \omega_i) N(R)}.
            \end{equation}
        This is a differential equation for $f(R)$, where the scalar curvature is here $R = 3 G'(N) + 12 G(N)$.

        \subsubsection{EXAMPLE: ASYMPTOTICALLY TRANSIENT PHANTOM BEHAVIOR}

            Let us consider an evolution given by the following Hubble parameter:
                \begin{equation}\label{t7}
                    H^2(N) = H_0 \ln{\left( \frac{a}{a_0} \right)} + H_1 = H_0 N + H_1 = G(N),
                \end{equation}
            where $H_0$ and $H_1$ are positive constants. We can, in this case, achieve a  phantom behavior with the possibility to be asymptotically transient,
            without the presence of real matter. The present is actually a simplified example, but can be of use as a component part of a more elaborated model where,
            with a modified functionality, the transition can be reached at finite time (this is work in progress, the results of which will be
            presented in a future publication.) Indeed, from $R = 3 G'(N)+ 12 G(N)$, we find
                \begin{equation}\label{t8}
                    N = \frac{R - 3 H_0}{12 H_0} - \frac{H_1}{H_0}.
                \end{equation}
            Eq.~(\ref{t6}) takes the form (with $r$ being the curvature measured in terms of $H_0$, $r\equiv R /H_0$)
                \begin{equation}\label{t9}
                    0 = - 3  (r - 3 ) \frac{d^2 f(r)}{dr^2} + \left( \frac{r + 3 }{4}\right) \frac{d f(r)}{dr} - \frac{f(r)}{2},
                \end{equation}
and changing now the variable from $r$ to $x$, as $x = \frac{r - 3}{12}$, Eq.~(\ref{t9}) reduces to
                \begin{equation}\label{t10}
                    0 =  x \frac{d^2 F(x)}{dx^2} - \left( x + \frac{1}{2} \right) \frac{d F(x)}{dx} + 2 F(x),
                \end{equation}
            which is easily recognized as a degenerate hypergeometric equation, whose solutions are given by the Kummer's series $\Phi (a, b; z)$, the simplest one being
                \begin{equation}\label{t11}
                    f(r) = C \, \Phi \left( -2, - \frac{1}{2}; \frac{r - 3 }{12} \right) = C_1 \left( - \frac{1}{2} + r  - \frac{r^2}{18}  \right),
                \end{equation}
            where $C$ is a constant. As a consequence, with {\it this} $f(R)$ theory, the solution given by Eq.~(\ref{t11}), we can reproduce the phantom behavior without real matter given by Eq.~(\ref{t7}).

            Taking this into account, for (\ref{t7}), we have
                \begin{equation}\label{a1}
                    H(t) = \frac{H_0}{2} (t - t_0),
                \end{equation}
            and it turns out that, with this model, we have a contribution of an effective cosmological constant and another term which will produce an accelerating phase but, remarkably, without developing a future singularity, in spite of its phantom nature. Hence, the $f(R)$ gravity given by Eq.~(\ref{t11}) gives rise to a cosmological solution, with an asymptotically transient phantom behavior, which does not evolve into a future singularity. This property relies on the fact that the phantom behavior gets more and more mild with time (asymptotically disappears), at a rate that overcomes the one for the formation of the singularity.

Actually, there is another independent solution of Kummer's equation (\ref{t10}), the complete solution being:
\begin{equation}\label{t11c}
f(r) = C_1 \left( - \frac{1}{2} + r  - \frac{r^2}{18}  \right) \ + \ C_2 \ \left( \frac{r - 3}{12} \right)^{3/2} L_{1/2}^{(3/2)} \left( \frac{r - 3}{12} \right),
\end{equation}
where the second basic solution is a Laguerre $L$ function, which is well behaved but cannot be represented as a simple rational one. It is interesting to note that
this second function asymptotically behaves exactly in the same way as the first, for large negative curvature (e.g., as $R^2$, when $R \to -\infty$). For large positive one it explodes exponentially, as $R^{-3/2} \, e^{R/12}$ (again, $R$ in units of $H_0$).

    \subsection{GENERAL FORMULATION USING A SCALAR FIELD}

        In this section it will be shown how to construct an $f(R)$ model realizing a given cosmology, but using this time a different technique, which involves a scalar field \cite{3}. The final aim will be to apply this procedure to the novel case with a Yang-Mills term, what will be performed in Sect.~III. Here we summarize the basic tools necessary in order to understand the procedure and to make the present paper self-contained. We start from the action for modified gravity
            \begin{equation}\label{t12}
                S = \int d^4 x \sqrt{-g} (f(R) + \mathcal{L}_{matter}),
            \end{equation}
        which is equivalent to
            \begin{equation}\label{t13}
                S = \int d^4 x \sqrt{-g} (P(\phi) R + Q(\phi) + \mathcal{L}_{matter}).
            \end{equation}
        Here, $\mathcal{L}_{matter}$ is the matter Lagrangian density and $P$ and $Q$ are proper functions of the scalar field, $\phi$, which can be regarded as an auxiliary field, because there is no kinetic term depending on $\phi$ in the Lagrangian. By varying the action with respect to $\phi$, $                 0 = P'(\phi) R + Q'(\phi)$, which can be solved in terms of $\phi$, as $\phi = \phi (R)$.
        Substituting it into (\ref{t13}) and comparing with (\ref{t12}), one obtains
            \begin{equation}\label{t16}
                S = \int d^4 x \sqrt{-g} (f(R) + {\mathcal L}_{matter}), \qquad f(R) \equiv P(\phi(R)) R + Q(\phi(R)),
            \end{equation}
        and  varying the action with respect to the metric $g_{\mu \nu}$,
            \begin{equation}\label{t17}
                0 = - \frac{1}{2} g_{\mu \nu} (P(\phi) R + Q(\phi)) + R_{\mu \nu} P(\phi) + g_{\mu \nu} \nabla^2 P(\phi) - \nabla_\mu \nabla_\nu P(\phi) - \frac{1}{2} T_{\mu \nu}.
            \end{equation}
        The equations corresponding to the standard, spatially-flat FRW universe are
            \begin{equation}\label{t18}
                0 = - Q(\phi) - 6 H^2 P(\phi) - 6 H \frac{d P(\phi)}{dt} + \rho,
            \end{equation}
            \begin{equation}\label{t19}
                0 = Q(\phi) + \left( 4 \dot H + 6 H^2 \right) P(\phi) + 4 H \frac{d P(\phi)}{dt} + 2 \frac{d^2 P(\phi)}{dt^2} + p,
            \end{equation}
        and, by combining them, we find
            \begin{equation}\label{t20}
                0 = 2 \frac{d^2 P(\phi(t))}{dt^2} - 2 H \frac{d P(\phi(t))}{dt} + 4 \dot H P(\phi(t)) + p + \rho.
            \end{equation}
        As we are allowed to redefine the scalar field $\phi$ properly, we choose the  most simple, non-constant, smooth possibility (what is commonly done in this kind of approaches), namely $\phi = t$.

Now, given a cosmology, specified through the scale factor $a$, given by a proper function $g(t)$ as
            \begin{equation}\label{t22}
                a = a_0 e^{g(t)},
            \end{equation}
        with a constant $a_0$, and if it is assumed that $p$ and $\rho$ consist of the sum of different matter contributions, each one with constant EoS parameter, $\omega_i$, then Eq.~(\ref{t20}) reduces to the following second order differential equation
            \begin{equation}\label{t23}
                0 = 2 \frac{d^2 P(\phi)}{d\phi^2} - 2 g'(\phi) \frac{d P(\phi)}{d\phi} + 4 g''(\phi) P(\phi) + \sum \limits_i (1 + \omega_i) \rho_{i0} a_0^{-3 (1 + \omega_i)} e^{-3 (1 + \omega_i) g(\phi)},
            \end{equation}
        from where one can obtain $P(\phi)$ and, using Eq.~(\ref{t18}),
            \begin{equation}\label{t24}
                Q(\phi) = - 6 (g'(\phi))^2 P(\phi) - 6 g'(\phi) \frac{d P(\phi)}{d\phi} + \sum \limits_i \rho_{i0} a_0^{-3 (1 + \omega_i)} e^{-3 (1 + \omega_i) g(\phi)}.
            \end{equation}
        As a result, and as anticipated, any given cosmology (\ref{t22}) can indeed be realized through some corresponding $f(R)$-gravity. Before applying this procedure to the Yang-Mills case, let us make things even more clear by means of the example considered before.

        \subsubsection{EXAMPLE: ASYMPTOTICALLY TRANSIENT PHANTOM BEHAVIOR}
            In order to compare the two different methods developed for the reconstruction of $f(R)$ gravities---to reproduce any given cosmology---we consider again the asymptotically transient phantom behavior, without real matter, given by (\ref{t7}). The Hubble parameter can be written as
                \begin{equation}\label{t25}
                    H = \sqrt{H_0 g(t) + H_1} = \frac{d g(t)}{dt},
                \end{equation}
            and thus
                \begin{equation}\label{t26}
                    g(t) = \frac{H_0}{4} (t - c)^2 - \frac{H_1}{H_0},
                \end{equation}
            with $c$ an integration constant. Introducing (\ref{t26}) into (\ref{t23}),
                \begin{equation}\label{t27}
                    0 = \frac{d^2 P(\phi)}{d\phi^2} - \frac{H_0}{2} (\phi - c) \frac{d P(\phi)}{d\phi} + H_0 P(\phi),
                \end{equation}
            and using a new variable, $x = \phi - c$, we get
                \begin{equation}\label{t28}
                    0 = \frac{d^2 P(x)}{dx^2} - \frac{H_0}{2} x \frac{d P(x)}{dx} + H_0 P(x),
                \end{equation}
            whose solution is
                \begin{equation}\label{t29}
                    P(x) = \frac{1}{2} \, (2 - H_0 x^2) \, C_1 \, + \, \frac{1}{2} \, (2 - H_0 x^2) \, C_2 \, \left( \frac{e^{\frac{H_0}{4} x^2} x}{4 (2 - H_0 x^2)} \, - \, \frac{i}{4 \sqrt{H_0}} \, \int \limits_0^{i \sqrt{\frac{H_0 x^2}{4}}} e^{-y^2} dy \right).
                \end{equation}
            Now, using (\ref{t18}), we obtain
                \begin{equation}\label{t30}
                    Q(x) = \frac{3}{32} H_0 x \left[ 8 H_0 x \left( 2 + H_0 x^2 \right) C_1 - \left( \left( 8 + 2 H_0 x^2 \right) e^{\frac{H_0 x^2}{4}} + i \, 2 \sqrt{H_0} x \left( 2 + H_0 x^2 \right) \int \limits_0^{i \sqrt{\frac{H_0 x^2}{4}}} e^{- y^2} dy \right) C_2 \right].
                \end{equation}
            Taking into account that $R = 6 \dot H + 12 H^2 = 6 g''(x) + 12 \left( g'(x) \right)^2$, it follows that
                \begin{equation}\label{t31}
                    x = \sqrt{\frac{R - 3 H_0}{3 H_0^2}}.
                \end{equation}
            Introducing at this step (\ref{t29}) and (\ref{t30}) into (\ref{t16}), and considering (\ref{t31}), one finally gets the explicit expression
                \begin{equation}\label{t32}
                    f(R) = - \frac{R^2 - 18 H_0 R + 9 H_0^2}{12 H_0} \, C_1 \, - \, \left[ e^{\frac{R - 3 H_0}{12 H_0}} \, \frac{R - 9 H_0}{16} - \frac{i}{24 H_0} \left( R^2 - 18 H_0 R + 9 H_0^2 \right) \int \limits_0^{i \sqrt{\frac{R - 3 H_0}{12 H_0}}} e^{- y^2} dy \right] \, C_2.
                \end{equation}
            We thus have proven that, within this scheme, we are able to obtain the $f(R)$ model (\ref{t32}) which reproduces the desired transient phantom behavior without real matter, as given by (\ref{t7}).

\section{RECONSTRUCTION OF THE YANG-MILLS THEORY}

    \subsection{GENERAL FORMULATION}

        In this section we develop a reconstruction scheme of the YM theory without any auxiliary scalar field. Consider the following action
            \begin{equation}\label{1}
                S = \int dx^4 \sqrt{-g} \left[ \frac{R}{2 \kappa^2} + \mathcal F \left( F_{\mu \nu}^a F^{a \, \mu \nu} \right) \right],
            \end{equation}
        where $F_{\mu \nu}^a = \partial_\mu A^a_\nu - \partial_\nu A^a_\mu + f^{abc} A^b_\mu A^c_\nu$, and $\mathcal F$ may be assumed to be a smooth function (however, strictly speaking it may suffice if it is continuously differentiable). The presence of this function is necessary in order to allow for more freedom in the choice of the theory \cite{25} (since difficulties inherent to the problem may prevent obtaining the standard Yang-Mills case). Anyway, this function will not constitute a problem for the development of our methods,
        which are thus proven to be even more powerful. For simplicity of the derivation (and in order not to break the line of argument of this paper)
        we concentrate here on the $SU(2)$ case where $f^{abc}=\epsilon^{abc}$ but, with some more effort,
        exactly the same procedure can be extended to other gauge groups (more general cases will be treated
        in a subsequent publication). Taking into account that
            \begin{equation}\label{2}
                \frac{\delta \left( F_{\mu \nu}^a F^{a \, \mu \nu} \right)}{\delta A_\beta^h} = - 4 \epsilon^{hbc} A^b_\gamma F^{c \, \gamma \beta},
            \end{equation}
            \begin{equation}\label{3}
                \frac{\delta \left( F_{\mu \nu}^a F^{a \, \mu \nu} \right)}{\delta \left( \partial_\alpha A_\beta^h \right)} = 4 F^{h \, \alpha \beta},
            \end{equation}
        the equation of motion for the field potential $A_\mu^a$ turns into
            \begin{equation}\label{4}
                \partial_\nu \left[ \frac{\delta S}{\delta \left( \partial_\nu A^a_\mu \right)} \right] - \frac{\delta S}{\delta A^a_\mu} = 0
            \end{equation}
        and, from here,
            \begin{equation}
                \partial_\nu \left[ \sqrt{-g} \, \mathcal F\,' \left( F^a_{\alpha \beta} F^{a \, \alpha \beta} \right) \, F^{a \, \nu \mu} \right] + \sqrt{-g} \, \mathcal F\,' \left( F^a_{\alpha \beta} F^{a \, \alpha \beta} \right) \, \epsilon^{abc} A^b_\nu F^{c \, \nu \mu} = 0.
            \end{equation}
        Variation of (\ref{1}) with respect to $g^{\mu \nu}$ yields the following equation of motion
            \begin{equation}\label{5}
                \frac{1}{2 \kappa^2} \left( R_{\mu \nu} - \frac{1}{2} g_{\mu \nu} R \right) - \frac{1}{2} g_{\mu \nu} \mathcal F \left( F^a_{\alpha \beta} F^{a \, \alpha \beta} \right) + 2 \mathcal F\,' \left( F^a_{\alpha \beta} F^{a \, \alpha \beta} \right) \, F^a_{\mu \rho} F^{a \, \rho}_\nu = 0,
            \end{equation}
        where we have used $\frac{\delta \left( F^a_{\rho \sigma} F^{a \, \rho \sigma} \right)}{\delta g^{\mu \nu}} = 2 \mathcal F\,' \left( F^a_{\alpha \beta} F^{a \, \alpha \beta} \right) \, F^a_{\mu \gamma} F^{a \, \gamma}_\nu$.
        Considering now a FRW universe, and the following Ansatz for the gauge field \footnote{Note that our aim here is to demonstrate that the procedure works, and not to find the most general solution which exhausts all possibilities. This issue, which is certainly interesting and more difficult, will be left to further consideration.},
            \begin{equation}\label{gauge}
                A_\mu^a = \left\{
                \begin{array}{ll}
                    \bar \alpha e^{\lambda (t)} \delta^a_\mu, & \mu = i, \\
                    0, & \mu = 0,
                \end{array}
                \right.
            \end{equation}
        the $\mu = 0$ component of (\ref{4}) becomes an identity, the $\mu = i$ component yields
            \begin{equation}\label{6}
                \partial_t \left[ a(t) \, \mathcal F\,' \left( F^a_{\alpha \beta} F^{a \, \alpha \beta} \right) \, \dot \lambda (t) \, e^{\lambda (t)} \right] + \frac{2 \bar \alpha^2}{a(t)} \, \mathcal F\,' \left( F^a_{\alpha \beta} F^{a \, \alpha \beta} \right) \, e^{3 \lambda(t)} = 0,
            \end{equation}
        while the $(t,t)$ component of (\ref{5}) is
            \begin{equation}\label{7}
                \frac{3 H^2(t)}{2 \kappa^2} + \frac{1}{2} \mathcal F \left( F^a_{\alpha \beta} F^{a \, \alpha \beta} \right) + 2 \bar \alpha^2 \, \mathcal F\,' \left( F^a_{\alpha \beta} F^{a \, \alpha \beta} \right) \, \frac{\dot \lambda^2 (t) e^{2 \lambda(t)}}{a^2(t)} = 0,
            \end{equation}
        and the $(i,i)$ component of (\ref{5}) reduces to
            \begin{equation}\label{8}
                \frac{1}{2 \kappa^2} \left[ 2 \dot H(t) + 3 H^2(t) \right] - \frac{1}{2} \mathcal F \left( F^a_{\alpha \beta} F^{a \, \alpha \beta} \right) - 2 \bar \alpha^2 \frac{e^{2 \lambda (t)}}{a^2 (t)} \mathcal F\,' \left( F^a_{\alpha \beta} F^{a \, \alpha \beta} \right) \left[ \dot \lambda^2 (t) - 2 \bar \alpha^2 \frac{e^{2 \lambda (t)}}{a^2 (t)} \right] = 0.
            \end{equation}
        Adding (\ref{7}) to (\ref{8}), one arrives at
            \begin{equation}\label{9}
                a^4 (t) \dot H (t) - 4 \kappa^2 \bar \alpha^4 \, \mathcal F\,' \left( F^a_{\alpha \beta} F^{a \, \alpha \beta} \right) \, e^{4 \lambda (t)} = 0,
            \end{equation}
        and  then
            \begin{equation}\label{10}
                \mathcal F\,' \left( F^a_{\alpha \beta} F^{a \, \alpha \beta} \right) = \frac{a^4 (t) \, \dot H (t)}{4 \kappa^2 \bar  \alpha^4} e^{-4 \lambda (t)}.
            \end{equation}
        Using (\ref{10}), Eq.~(\ref{6}) reduces to:
            \begin{equation}\label{11}
                2 \bar \alpha^2 \dot H (t) \, e^{2 \lambda (t)} \, + \, \left[ 5 a(t) \, \dot a(t) \, \dot H(t) \, + \, a^2(t) \, \ddot H(t) \right] \dot \lambda (t) \, - \, 3 a^2(t) \dot H(t) \dot \lambda^2 (t) \, + \, a^2(t) \dot H(t) \ddot \lambda (t) = 0,
            \end{equation}
        which constitutes a differential equation for $\lambda (t)$. Hence, by using Eq.~(\ref{gauge}), once we have the function $\lambda (t)$, given by (\ref{11}), we can obtain the corresponding YM theory which reproduces the selected cosmology. The Ansatz considered above actually leads to a mathematical solution of the problem.

        \subsubsection{EXAMPLE: POWER LAW EXPANSION}

            Considering the case of power law expansion: $a(t) = \left( \frac{t}{t_1} \right)^{h_1}$, where $t_1$ and $h_1$ are constant, and assuming $\lambda (t) = (h_1 - 1) \ln{\left( \frac{t}{t_1} \right)} + \lambda_1$, where $\lambda_1$ is again a constant, Eq.~(\ref{11})  reduces to the following algebraic equation
                \begin{equation}\label{12}
                    h_1 (h_1 - 1) + \bar \alpha^2 t_1^2 \, e^{2 \lambda_1} = 0,
                \end{equation}
            hence
                \begin{equation}\label{13}
                    \lambda_1 = \frac{1}{2} \ln{\left( \frac{h_1 (1 - h_1)}{\bar \alpha^2 t_1^2} \right)}
                \end{equation}
            and
                \begin{equation}\label{13b}
                    \lambda (t) = (h_1 - 1) \ln{\left( \frac{t}{t_1} \right)} + \frac{1}{2} \ln{\left( \frac{h_1 (1 - h_1)}{\bar \alpha^2 t_1^2} \right)}.
                \end{equation}
            With the help of this reconstruction scheme, we have obtained the function $\lambda (t)$ given by (\ref{13b}). Then, using (\ref{gauge}), we are able to reproduce the cosmology given by the power law expansion: $a(t) = \left( \frac{t}{t_1} \right)^{h_1}$.

    \subsection{GENERAL FORMULATION USING A SCALAR FIELD}
        In this section, the reconstruction of a YM theory is performed using the technique of \cite{26}. By way of introducing an auxiliary scalar field, $\phi$, we can rewrite action (\ref{1}) as
            \begin{equation}\label{14}
                S = \int d^4 x \sqrt{-g} \left( \frac{R}{2 \kappa^2} + \frac{1}{4} P(\phi) F^a_{\mu \nu} F^{a \, \mu \nu} + \frac{1}{4} Q(\phi) \right).
            \end{equation}
        Variation of (\ref{14}) with respect to $\phi$ yields the corresponding  equation of motion
            \begin{equation}\label{15}
                0 = F^a_{\mu \nu} F^{a \, \mu \nu} \frac{d P(\phi)}{d \phi} + \frac{d Q(\phi)}{d \phi},
            \end{equation}
        which can be solved with respect to $\phi$ as $\phi = \phi (F^a_{\mu \nu} F^{a \, \mu \nu})$. Then, action (\ref{14}) is rewritten as
            \begin{equation}\label{16}
                S = \int dx^4 \sqrt{-g} \left[ \frac{R}{2 \kappa^2} + \mathcal F \left( F_{\mu \nu}^a F^{a \, \mu \nu} \right) \right],
            \end{equation}
        where
            \begin{equation}\label{17}
                \mathcal F \left( F_{\mu \nu}^a F^{a \, \mu \nu} \right) = \frac{1}{4} P(\phi (F^a_{\mu \nu} F^{a \, \mu \nu})) F^a_{\mu \nu} F^{a \, \mu \nu} + \frac{1}{4} Q(\phi (F^a_{\mu \nu} F^{a \, \mu \nu})).
            \end{equation}
        Taking the variations of this action (\ref{14}) with respect to $g_{\mu \nu}$, we obtain the Einstein equation
            \begin{equation}\label{18}
                \frac{1}{2 \kappa^2} \left( R_{\mu \nu} - \frac{1}{2} g_{\mu \nu} R \right) = - \frac{1}{2} P(\phi) F^a_{\mu \rho} F^{a \, \rho}_\nu + \frac{1}{8} g_{\mu \nu} \left( P(\phi) F^a_{\alpha \beta} F^{a \, \alpha \beta} + Q(\phi) \right).
            \end{equation}
        Finally, taking the variations of (\ref{14}) with respect to $A_\mu^a$, it follows that
            \begin{equation}\label{19}
                0 = \partial_\nu \left( \sqrt{-g} P(\phi) F^{a \, \nu \mu} \right) + \sqrt{-g} P(\phi) f^{abc} A^b_\nu F^{c \, \nu \mu}.
            \end{equation}
        We restrict our analysis to the case where the gauge algebra is $SU(2)$ and the gauge fields are given by
            \begin{equation}\label{20}
                A_\mu^a = \left\{
                    \begin{array}{ll}
                        \bar \alpha e^{\lambda (t)} \delta^a_\mu, & \mu = i, \\
                        0, & \mu = 0.
                    \end{array}
                \right.
            \end{equation}
        With these assumptions, Eq.~(\ref{15}) reduces to
            \begin{equation}\label{21}
                0 = 6 \left( - \bar \alpha^2 \dot \lambda (t)^2 e^{2 \lambda (t)} a(t)^{-2} + \bar \alpha^4 e^{4 \lambda (t)} a(t)^{-4} \right) \frac{d P(\phi)}{d \phi} + \frac{d Q(\phi)}{d \phi}.
            \end{equation}
        The $(t,t)$ component of (\ref{18}) is
            \begin{equation}\label{22}
                0 = \frac{3 H(t)^2}{\kappa^2} + \frac{3}{2} \left( \bar \alpha^2 \dot \lambda (t)^2 e^{2 \lambda (t)} a(t)^{-2} + \bar \alpha^4 e^{4 \lambda (t)} a(t)^{-4} \right) P(\phi) + \frac{1}{4} Q(\phi),
            \end{equation}
        the $(t,i)$ component of (\ref{18}) becomes an identity, while the component $(i,j)$ is
            \begin{equation}\label{spatial component}
                \left[ -\frac{1}{2 \kappa^2} \left( 2 \dot H(t) + 3 H(t)^2 \right) \right] \delta_{ij} = \left[ - \frac{1}{4} P(\phi) \left( \bar \alpha^2 \dot \lambda(t)^2 e^{2 \lambda(t) a(t)^{-2} + \bar \alpha^4 e^{4 \lambda(t)} a(t)^{-4}} \right) + \frac{1}{8} Q(\phi) \right] \delta_{ij}.
            \end{equation}
        The $\mu = 0$ component of (\ref{19}) becomes an identity, and the $\mu = i$ component yields
            \begin{equation}\label{23}
                0 = \partial_t \left( a(t) P(\phi) \dot \lambda (t) e^{\lambda (t)} \right) + 2 \bar \alpha^2 a(t)^{-1} P(\phi) e^{3 \lambda (t)}.
            \end{equation}
        Here, we can identify $\phi = t$, because we are always allowed to take the scalar field $\phi$ properly in order to satisfy this. By differentiating (\ref{22}) with respect to $t$ and eliminating $\dot Q = \frac{d Q(\phi)}{d \phi}$, it follows that
            \begin{equation}\label{24}
                0 = \frac{2}{\kappa^2} H(t) \dot H(t) + \bar \alpha^2 \dot \lambda (t)^2 e^{2 \lambda (t)} a(t)^{-2} \dot P (t) +$$
                $$+ \left[ \bar \alpha^2 \left( \dot \lambda (t) \ddot \lambda (t) + \dot \lambda (t)^3 - \dot \lambda (t)^2 H(t) \right) e^{2 \lambda (t)} a(t)^{-2} + 2 \bar \alpha^4 \left( \dot \lambda (t) - H(t) \right) e^{4 \lambda (t)} a(t)^{-4} \right] P(t).
            \end{equation}
        Using (\ref{23}), we can solve for $\dot P$ in (\ref{24}), and  obtain
            \begin{equation}\label{25}
                P = \frac{a(t)^2 \dot H(t)}{\kappa^2 \bar \alpha^2 e^{2 \lambda (t)} \left[ \dot \lambda (t)^2 + \bar \alpha^2 e^{2 \lambda (t)} a(t)^{-2} \right]}.
            \end{equation}
        Taking into account (\ref{25}), Eq.~(\ref{23}) reduces to
            \begin{equation}\label{26}
                0 = 2 \dot H(t) \left( \bar \alpha^2 e^{2 \lambda (t)} a(t)^{-2} \right)^2 + \bar \alpha^2 e^{2 \lambda (t)} a(t)^{-2} \left[ \dot \lambda (t) \left( 5 H(t) \dot H(t) + \ddot H(t) \right) - \dot \lambda (t)^2 \dot H(t) + \ddot \lambda (t) \dot H(t) \right] +$$
                $$+ \dot \lambda (t)^2 \left[ \dot \lambda (t) \left( 3 H(t) \dot H(t) + \ddot H(t) \right) - \dot \lambda (t)^2 \dot H(t) - \ddot \lambda (t) \dot H(t) \right],
            \end{equation}
        which constitutes a differential equation for $\lambda (t)$. As was the case for the other reconstruction scheme, developed in section III.A, once we have the function $\lambda (t)$---given here by (\ref{26})---we can readily obtain the modified YM theory that reproduces the desired cosmology, through the use of Eq.~(\ref{20}), which was our starting Ansatz. Note that in (\ref{24}) we have positively corrected some missprints of a previous calculation (recognized by the authors).
        Using Eq. (\ref{22}) and (\ref{25}), it is easy to check that Eq.~(\ref{spatial component}) becomes an identity, thus being always fulfilled.

        \subsubsection{EXAMPLE: POWER LAW EXPANSION}
            Consider now the case of power law expansion: $a(t) = \left( \frac{t}{t_1} \right)^{h_1}$, where $t_1$ and $h_1$ are constants and assume $\lambda (t) = (h_1 - 1) \ln{\left( \frac{t}{t_1} \right)} + \lambda_1$, where $\lambda_1$ is a constant. Eq.~(\ref{26}) turns into the algebraic one
                \begin{equation}\label{27}
                    0 = \bar \alpha^4 t_1^4 e^{4 \lambda_1} + \bar \alpha^2 t_1^2 e^{2 \lambda_1} (h_1 - 1) (2 h_1 - 1) + h_1 (h_1 - 1)^3,
                \end{equation}
            hence
                \begin{equation}\label{28}
                    \lambda_1 = \left\{
                        \begin{array}{lll}
                            \frac{1}{2} \ln{\left( \frac{(h_1 - 1) (1 - h_1)}{\bar \alpha^2 t_1^2} \right)}, \\
                            \\ \frac{1}{2} \ln{\left( \frac{h_1 (1 - h_1)}{\bar \alpha^2 t_1^2} \right)},
                        \end{array}
                    \right.
                \end{equation}
            and then
                \begin{equation}\label{28b}
                    \lambda (t) = (h_1 - 1) \ln{\left( \frac{t}{t_1} \right)} + \left\{
                        \begin{array}{lll}
                            \frac{1}{2} \ln{\left( \frac{(h_1 - 1) (1 - h_1)}{\bar \alpha^2 t_1^2} \right)}, \\
                            \\ \frac{1}{2} \ln{\left( \frac{h_1 (1 - h_1)}{\bar \alpha^2 t_1^2} \right)}.
                        \end{array}
                    \right.
                \end{equation}
            With the function $\lambda (t)$ given by (\ref{28b}) and using (\ref{20}), we can now reproduce the cosmology given by the power law expansion: $a(t) = \left( \frac{t}{t_1} \right)^{h_1}$.

\section{SUMMARY AND DISCUSSION}

    In this paper, after carefully reviewing and comparing (for the first time), and also with the help of corresponding examples, two different schemes of reconstructing cosmologies for modified gravity, we have successfully extended them to the case of YM theories. The first scheme does not need an auxiliary scalar field, while the second one is thoroughly based on its use. With these reconstruction methods, any explicitly given cosmology can be realized as a corresponding modified gravity or YM theory, respectively. Although this fact had been already anticipated in the specialized literature---for the first of the two situations and concerning some basic models---it is comforting to see here how it can be also explicitly extended to more realistic physical theories, as the modified Yang-Mills one, with reasonable effort. As we have indicated in the paper, things are far from straightforward, and a very careful analysis of the solutions here obtained (also in relation with the comparison of the two different methods), and of other additional solutions of the differential equations, with potential physical interest, is still necessary. First results indicate that the solutions obtained do pass the solar system tests and the other known physical constraints.

    In order to compare both schemes of reconstruction, an example has been explicitly worked out in the two cases. The result obtained in the first scheme (in terms of e-foldings) is
        \begin{equation}\label{comp1}
            f(R) = C \, \Phi \left( -2, - \frac{1}{2}; \frac{R - 3 H_0}{12 H_0} \right) = C \left( - \frac{1}{4} + \frac{1}{2 H_0} R - \frac{1}{36 H_0^2} R^2 \right),
        \end{equation}
while for the second scheme (using an auxiliary scalar field), the $f(R)$ obtained has the form
        \begin{equation}\label{comp2}
            f(R) = - \frac{R^2 - 18 H_0 R + 9 H_0^2}{12 H_0} \, C_1 \, - \, \left[ e^{\frac{R - 3 H_0}{12 H_0}} \, \frac{R - 9 H_0}{16} - \frac{i}{24 H_0} \left( R^2 - 18 H_0 R + 9 H_0^2 \right) \int \limits_0^{i \sqrt{\frac{R - 3 H_0}{12 H_0}}} e^{- y^2} dy \right] \, C_2.
        \end{equation}
    As one can easily see, the results obtained for both methods are in fact different. The reason behind this is the fact that action (\ref{t13}) corresponds to a wider class of theories than action (\ref{t12}) (for a related and quite detailed discussion, see \cite{27,28}). Nevertheless, if in Eq.~(\ref{comp2}) we set $C_2 = 0$, then the results coming from both schemes are similar, at least in the sense that, for low curvatures, they behave as constant, while for large curvatures the behavior is in both cases proportional to $R^2$.

For the novel case of the reconstruction of a YM theory, we followed the same procedure, by considering also the same example for both reconstruction schemes. The first one yields the result
        \begin{equation}\label{comp3}
            \lambda_1 = \frac{1}{2} \ln{\left( \frac{h_1 (1 - h_1)}{\bar \alpha^2 t_1^2} \right)},
        \end{equation}
    while the second scheme gave the following one
        \begin{equation}\label{comp4}
            \lambda_1 = \left\{
                \begin{array}{lll}
                    \frac{1}{2} \ln{\left( \frac{(h_1 - 1) (1 - h_1)}{\bar \alpha^2 t_1^2} \right)}, \\
                    \\ \frac{1}{2} \ln{\left( \frac{h_1 (1 - h_1)}{\bar \alpha^2 t_1^2} \right)}.
                \end{array}
            \right.
        \end{equation}
As before, in the new situation considered in this paper of reconstructing a YM theory, it also happens that action (\ref{14}) expresses a more extensive class of theories than the action given by (\ref{1}), and it is again for this reason that more solutions are obtained for the scheme based on an auxiliary scalar field. Moreover, with the help of this example we could see explicitly that there is in fact a very interesting coincidence between the result obtained in (\ref{comp3}) and one of the results of (\ref{comp4}). This finding here further supports the point of view that the Einsteinian and the Jordanian frame descriptions actually lead to two physically different theories, making thus clear the physical non-equivalence of the two frames as discussed in the first and in the third references of \cite{23}. In view of the strong and still on going discussion about this issue in the specialized literature, this additional piece of evidence is very valuable. Even much more because it comes from a theory that it is way closer to physics than the ones considered previously.

Also important is to remark that, sometimes, it is actually more convenient to use one scheme instead of the other, because the final result---after following the way to find the modified or non-minimal gravity that reproduces a given desired cosmology---may be definitely easier to obtain and to interpret in one of the two schemes. To repeat, although these conclusions may not seem really new, since they were already derived in more simplified situations, it is actually comforting (and rather non-trivial) to see that they continue to be valid in much more realistic situations, from the point of view of physics, as corresponding to the actions here considered, involving Yang-Mills fields.

        \begin{acknowledgements}
We are indebted with the referee for a careful revision of the manuscript that lead to significant improvement. We thank Sergei Odintsov for suggesting the problem and giving ideas to carry out this task. The work has been supported by MICINN (Spain), project FIS2006-02842, and by AGAUR (Generalitat de Ca\-ta\-lu\-nya), contract 2009SGR-994.
        \end{acknowledgements}


\begin{thebibliography}{50}
    \bibitem{4}
        S. Perlmutter \textit{et al.}, Astrophys. J. \textbf{517}, 565 (1999).
    \bibitem{5}
        A. G. Riess \textit{et al.}, Astron. J. \textbf{116}, 1009
(1998); Astron. J. \textbf{117}, 707 (1999).
\bibitem{eli1} E. Elizalde,
J. Phys. {\bf A39}, 6299 (2006) [arXiv:hep-th/0607185].
    \bibitem{8}
        H.K. Jassal, J.S. Bagla and T. Padmanabhan, Phys.\ Rev.\  D {\bf 72}
(2005) 103503 [arXiv:astro-ph/0506748];
[arXiv:astro-ph/0601389]; S. Nesseris and L. Perivolaropoulos, Phys.\ Rev.\  D {\bf 73}
(2006) 103511 [arXiv:astro-
ph/0602053]; [arXiv:astro-ph/0610092].
    \bibitem{9}
        M. Majumdar and A. C. Davis, Phys.\ Rev.\  D {\bf 69} (2004) 103504.
    \bibitem{1}
        S. Nojiri and S. D. Odintsov, eConf \textbf{C0602061}, 06 (2006)
[Int. J. Geom. Meth. Mod. Phys. \textbf{4}, 115 (2007)] [arXiv:hep-
th/0601213]; [arXiv:0807.0685]; S. Capozziello and M. Francaviglia, Gen. Rel.
Grav. \textbf{40}, 357 (2008) [arXiv:astro-ph/0706.1146]; T. P. Sotiriou and
V. Faraoni, [arXiv:gr-qc/0805.1726]; F. S. N. Lobo, [arXiv:gr-qc/0807.1640].
    \bibitem{16}
        S.~Nojiri and S.~D.~Odintsov, Phys.\ Rev.\ D {\bf 68}, (2003)
123512 [arXiv:hep-th/0307288].
    \bibitem{10}
        S. Capozziello, Int. J. Mod. Phys. D {\bf 11} (2002) 483; S.
Capozziello, S. Carloni and A. Troisi, Recent \ Res.\ Dev.\ Astron.\ Astrophys.\ {\bf 1} 625 (2003) [arXiv:astro-ph/0303041].
    \bibitem{11}
        S.M. Carroll, V. Duvvuri, M. Trodden and M.S. Turner, Phys.\ Rev.\
D {\bf 70} (2004) 043528.
    \bibitem{12}
        S. Nojiri and S. D. Odintsov, Phys.\ Lett.\  B {\bf 576} (2003)
5, [arXiv:hep-th/0307071].
    \bibitem{13}
        S. Nojiri, S. D. Odintsov and M. Sasaki, Phys.\ Rev.\  D {\bf 71}
(2005) 123509, [arXiv:hep-th/0504052].

\bibitem{kom1} E. Komatsu et al. (WMAP Collaboration), Astrophys. J. Suppl.
Ser. {\bf 180}, 330 (2009); D.N. Spergel et al. (WMAP Collaboration), ibid.
{\bf 170}, 377 (2007).
    \bibitem{7}
        S. Nojiri and S. D. Odintsov, Phys.\ Lett.\  B {\bf 657}, 238
(2007), [arXiv:hep-th/0707.1941]; S. Nojiri and S. D. Odintsov, Phys.\ Rev.\
D {\bf 77}, 026007 (2008), [arXiv:hep-th/0710.1738]; G. Cognola, E. Elizalde,
S. Nojiri, S.D. Odintsov, L. Sebastiani and S. Zerbini, Phys.\ Rev.\  D {\bf
77}, 046009 (2008), [arXiv:hep-th/0712.4017].
    \bibitem{14}
        S. Capozziello, V. F. Cardone and A. Troisi, Phys.\ Rev.\  D {\bf
71} (2005) 043503; S. Capozziello, V. F. Cardone and M. Francaviglia, GRG
{\bf 38} (2006) 711; M. Amarzguioui, O. Elgaroy, D. F. Mota and T. Multamaki, Astron.\ Astrophys.\ {\bf 454} 707-714 (2006)
[arXiv:astro-ph/0510519]; O. Mena, J. Santiago and J. Weller, Phys. Rev.
Lett. {\bf 96} (2006) 041103; T. Koivisto and H. Kurki-Suonio, Class. Quant.
Grav. {\bf 23} (2006) 2355; S. Capozziello, V. F. Cardone, E. Elizalde, S.
Nojiri and S. D. Odintsov, Phys.\ Rev.\  D {\bf 73} (2006) 043512,
[arXiv:astro-ph/0508350]; N.J. Poplawski, Phys.\ Rev.\  D {\bf 74} (2006) 084032 [arXiv:gr-qc 0607124]; A. Borowiec,
W. Godlowski and M. Szydlowski, Phys.\ Rev.\  D {\bf 74} (2006) 043502 [arXiv:astro-ph/0602526]; S. Capozziello, A.
Stabile and A. Troisi, Mod.\ Phys.\ Lett.\ A{\bf 21} 2291-2301 (2006) [arXiv:gr-qc/0603071]; G. Allemandi and M. Ruggiero, Gen.\ Rel.\ Grav.\ {\bf 39} 1381 (2007)
[arXiv:astro-ph/0610661]; Y.S. Song, W. Hu and I. Sawicki, Phys.\ Rev.\  D {\bf 75} (2007) 044004 [arXiv:astro-
ph/0610532]; B. Li, K.C. Chan and M.C. Chu, Phys.\ Rev.\  D {\bf 76} (2007) 024002 [arXiv:astro-ph/0610794]; S. Mendoza
and Y. Rosas-Guevara, [arXiv:astro-ph/0610390].
    \bibitem{15}
        A. W. Brookfield, C. van de Bruck and L.M.H. Hall, Phys. Rev. D {\bf 74}, 064028 (2006) [arXiv:hep-th/0608015].
    \bibitem{17}
        G.~Allemandi, M.~Francaviglia, M.~Ruggiero and A.~Tartaglia, GRG {\bf 37} (2005) 1891
[arXiv:gr-qc/0506123]; S.~Capozziello, [arXiv:gr-qc/0412088]; X.~Meng and
P.~Wang, GRG {\bf 36} (2004) 1947; A.E.~Dominguez and D.E.~Barraco, Phys.\ Rev.\
D {\bf 70} (2004) 043505; T.~Koivisto, Class.\ Quant.\ Grav. {\bf 23} (2006) 4289 [arXiv:gr-qc/0505128]; T.~Clifton and
J.D.~Barrow, Phys.\ Rev.\ D {\bf 72} (2005) 103005; J.A.R.~Cembranos, Phys.\ Rev.\
D {\bf 73} (2006) 064029; T.~Sotiriou, GRG {\bf 38} (2006) 1407 [arXiv:gr-qc/0507027]; C.~Shao,
R.~Cai, B.~Wang and R.~Su, Phys.\ Lett.\ B {\bf 633} (2006) 164 [arXiv:gr-qc/0511034]; M.~E.~Soussa and
R.~P.~Woodard, GRG {\bf 36} (2004) 855; S.~Capozziello and A.~Troisi, Phys.\
Rev.\ D {\bf 72} (2005) 044022; K.~Atazadeh and H.~Sepangi, Int. J. Mod. Phys. D {\bf 16} (2007) 687 [arXiv:gr-
qc/0602028]; R.~Woodard, Lect.\ Notes Phys. {\bf 720} (2007) 403 [arXiv:astro-ph/0601672]; X.~Jin, D.~Liu and X.~Li,
[arXiv:astro-ph/0610854]; S. Nojiri and S. D. Odintsov, GRG {\bf 36} (2004)
1765, [arXiv:hep-th/0308176].
    \bibitem{18}
        A.~W.~Brookfield, C.~van~de~Bruck and L.M.H.~Hall, Phys.\ Rev.\ D {\bf 74} (2006) 064028 [arXiv:hep-th/0608015].
    \bibitem{19}
        A.~D.~Dolgov and M.~Kawasaki, Phys.\ Lett.\ B {\bf 573} (2003) 1;
V.~Faraoni, Phys.\ Rev.\ D {\bf 74} (2006) 104017 [arXiv:astro-ph/0610734]; V.~Faraoni,
Phys.\ Rev.\ D {\bf 74} (2006) 023529 [arXiv:gr-qc/0607016].
    \bibitem{20}
        I.~Brevik, S.~Nojiri, S.~D.~Odintsov and L.~Vanzo, Phys.\ Rev.\
D{\bf 70} (2004) 043520 [arXiv:hep-th/0401073]; G.~Cognola, E.~Elizalde,
S.~Nojiri, S.~D.~Odintsov and S.~Zerbini, JCAP {\bf 0502} 010 [arXiv:hep-
th/0501096]; B.C.~Paul and D.~Paul, Phys.\ Rev.\ D {\bf 74} (2006) 084015 [arXiv:hep-th/0511003]; Y.~Sobouti,
[arXiv:astro-ph/0603302]; T.~Multamaki and I.~Vilja, Phys.\ Rev.\ D {\bf 74} (2006) 064022 [arXiv:astro-
ph/0606373].
    \bibitem{6}
        W. Hu and I. Sawicki, PRD {\bf 76} 064004 (2007); S. Nojiri and
S. D. Odintsov, Phys.\ Lett.\  B {\bf 652} (2007) 343, [arXiv:hep-
th/0706.1378].
    \bibitem{29}
        S. Nojiri, S. D. Odintsov and S. Tsujikawa, Phys.\ Rev.\ D {\bf
71}, 063004 (2005), [arXiv:hep-th/0501025].
    \bibitem{32}
        S. Nojiri and S. D. Odintsov, Phys.\ Rev.\  D {\bf 78}, 046006
(2008), [arXiv:hep-th/0804.3519].
    \bibitem{33}
        S. Capozziello, M. De Laurentis, S. Nojiri and S. D. Odintsov,
GRG {\bf 41} (2009) 2313.
    \bibitem{31}
        M.C.B. Abdalla, S. Nojiri and S. D. Odintsov, Class.\ Quant.\
Grav.\ {\bf 22}, L35 (2005), [arXiv:hep-th/0409177].
    \bibitem{30}
        S. Nojiri and S.D. Odintsov, [arXiv:hep-th/0910.1464].
    \bibitem{21}
        S. Nojiri, S.D. Odintsov and D. S\'aez-G\'omez, [arXiv:hep-th/0908.1269].
 \bibitem{relw} W. Zhao and \, Y. Zhang, Class. Q. Grav. {\bf 23}, 3405 (2006);
Phys. Lett. {\bf B640}, 69 (2006); W. Zhao,\, Y. Zhang and \, M.L. Tong, arXiv:0909.3874;
S. Baghram and S. Rahvar, Phys. Rev. {\bf D80}, 124049 (2009).
    \bibitem{22}
        S.~Nojiri and S.~D.~Odintsov, J.\ Phys.\ Conf.\ Ser.\  {\bf 66},
012005 (2007) [arXiv:hep-th/0611071].
    \bibitem{23}
        S.~Nojiri and S.~D.~Odintsov, Phys.\ Rev.\  D {\bf 74}, 086005
(2006) [arXiv:hep-th/0608008]; J.\ Phys.\ A  {\bf 40}, 6725 (2007)
[arXiv:hep-th/0610164]; S.~Capozziello, S.~Nojiri, S.~D.~Odintsov and
A.~Troisi, Phys.\ Lett.\  B {\bf 639}, 135 (2006) [arXiv:astro-ph/0604431];
E.~Elizalde and D.~S\'{a}ez-G\'{o}mez, Phys.\ Rev.\  D {\bf 80}, 044030
(2009) [arXiv:hep-th/0903.2732]; E. Elizalde,
R. Myrzakulov, V.V. Obukhov, and D. S\'{a}ez-G\'{o}mez, Class. Quantum Grav.
{\bf 27}, 095007 (2010) [arXiv:hep-th/1001.3636].
    \bibitem{24}
        A.~de la Cruz-Dombriz and A.~Dobado, Phys.\ Rev.\  D {\bf 74},
087501 (2006) [arXiv:gr-qc/0607118]; J.~L.~Cortes and J.~Indurain,
Astropart.\ Phys.\  {\bf 31}, 177 (2009) [arXiv:astro-ph/0805.3481];
I.~H.~Brevik, GRG  {\bf 38}, 1317 (2006) [arXiv:gr-
qc/0603025]; L.~N.~Granda, [arXiv:hep-th/0812.1596]; X.~Wu and Z.~H.~Zhu,
Phys.\ Lett.\  B {\bf 660}, 293 (2008) [arXiv:astro-ph/0712.3603].
    \bibitem{25}
        K.~Bamba, C.~Q.~Geng, S.~Nojiri and S.~D.~Odintsov, Phys.\ Rev.\
D {\bf 79}, 083014 (2009) [arXiv:0810.4296 [hep-th]]; K.~Bamba, S.~Nojiri and
S.~D.~Odintsov, JCAP {\bf 0810}, 045 (2008) [arXiv:0807.2575 [hep-th]];
K.~Bamba and C.~Q.~Geng, Phys.\ Lett.\  B {\bf 679}, 282 (2009) [arXiv:hep-th/0901.1509].
    \bibitem{2}
        S. Nojiri, S. D. Odintsov and D. S\'aez-G\'omez, [arXiv:hep-
th/0908.1269].
    \bibitem{3}
        S. Nojiri and S. D. Odintsov, Phys.\ Rev.\ D {\bf 74}, 086005 (2006) [arXiv:hep-th/0608008].
    \bibitem{26}
        K. Bamba, S. Nojiri and S. D. Odintsov, Phys.\ Rev.\ D {\bf 77}, 123532 (2008) [arXiv:hep-th/0803.3384].
    \bibitem{27}
        S. Nojiri, S. D. Odintsov, A. Toporensky and P. Tretyakov,
[arXiv:hep-th/0912.2488].
    \bibitem{28}
        J. M. Pons, {\it Substituting fields within the action: consistency issues and some applications} [arXiv:hep-th/0909.4151].

    \end{thebibliography}
\end{document}